\pdfoutput=1
\documentclass[preprint,showpacs,preprintnumbers,amsmath,amssymb,floatfix,endfloats*]{revtex4}

\usepackage{graphicx}
\usepackage{dcolumn}
\usepackage{bm}
\usepackage{amsfonts}

\DeclareMathAlphabet{\mathsfsl}{OT1}{cmr}{bx}{it}
\begin{document}
\title{Wetting properties of structured interfaces composed of surface-attached spherical nanoparticles}
\author{Bishal Bhattarai and Nikolai V. Priezjev}
\affiliation{Department of Mechanical and Materials Engineering,
Wright State University, Dayton, Ohio 45435}
\date{\today}
%
\begin{abstract}

The influence of the external pressure and surface energy on the
wetting transition at nanotextured interfaces is studied using
molecular dynamics and continuum simulations. The surface roughness
of the composite interface is introduced via an array of spherical
nanoparticles with controlled wettability.  We find that in the
absence of an external pressure, the liquid interface is flat and
its location relative to the solid substrate is determined by the
particle size and the local contact angle.  With increasing pressure
on the liquid film, the interface becomes more curved and the
three-phase contact line is displaced along the spherical surface
but remains stable due to re-entrant geometry. It is demonstrated
that the results of molecular dynamics simulations for the critical
pressure of the Cassie-Baxter wetting state agree well with the
estimate of the critical pressure obtained by numerical minimization
of the interfacial energy.

\end{abstract}

\pacs{68.08.-p, 66.20.-d, 83.10.Rs}


\maketitle

\section{Introduction}

Over the last two decades, there has been a remarkable progress in
designing and fabricating the so-called \textit{superhydrophobic}
surfaces, which are characterized by small-scale surface roughness
that keeps the liquid interface suspended at the tips of the
asperities, thus, reducing the liquid-solid contact
area~\cite{Ramiasa17,Calama07}. Such engineered surfaces typically
exhibit large liquid contact angles, small contact angle hysteresis,
low adhesion, and reduced hydrodynamic friction, which play
important roles in many technological processes, including
self-cleaning surfaces, such as glasses and
fabrics~\cite{Selfclean}, nonfouling surfaces~\cite{Fouling}, and
drag reduction~\cite{Dragred}. Superhydrophobic textures can also
enable a more accurate control and manipulation of liquid flows in
microfluidic and nanofluidic
systems~\cite{Vinograd12,Neto14,BocqScale07}.  Notably, it was shown
that laminar flows over anisotropic textured surfaces can be
generally described via the mobility tensor that relates the normal
traction at the interface and the effective slip
velocity~\cite{VinogradJFM}.   From a modeling perspective, a
detailed comparison between continuum predictions and atomistic
simulations demonstrated that there is excellent agreement between
the velocity profiles and the effective slip boundary conditions if
the length scales of surface patterns are large compared to the
liquid molecular size~\cite{Priezjev05,Priezjev11}.

\vskip 0.05in

It is well known that, depending on the surface energy and
topography, a liquid droplet in contact with a roughened substrate
can form either the Wenzel state~\cite{Wenzel36}, where the liquid
fully penetrates into the surface asperities, or the Cassie-Baxter
state~\cite{CB44}, where the liquid interface remains suspended at
the tips of surface protrusions and small pockets of air become
trapped between the surface and the liquid.   In the latter case,
the \textit{apparent} contact angle is given by the Cassie-Baxter
equation:
\begin{equation}
\textrm{cos}\,\theta_c = (1-f)\,\textrm{cos}\,\theta - f,
\label{eq:cassie}
\end{equation}
where $f$ is the areal fraction of the liquid-air interface and
$\theta$ is the \textit{intrinsic} contact angle of a liquid droplet
at a smooth surface of the same material.   In particular, it
follows from Eq.\,(\ref{eq:cassie}) that in the limit when the areal
fraction $f$ is close to 1, the apparent contact angle becomes
nearly $180^{\circ}$.   In practice, however, the distance between
surface protrusions should be sufficiently small, which allows the
liquid-air interface to remain locally suspended above the surface;
otherwise, the liquid would fully wet the substrate.    Thus, the
main factors that limit the applicability of superhydrophobic
surfaces include the breakthrough pressure required to fully wet the
substrate, fluid temperature, fragility of the surface texture, and
fouling resistance~\cite{Ramiasa17}.

\vskip 0.05in

More generally, the types of surfaces that exhibit highly repellent
properties for liquids with both high and low surface tension (for
example, water and oil) are called \textit{omniphobic}
surfaces~\cite{McKinley07,McKinley08}.   Such an unusual property
can be achieved by introducing the re-entrant surface curvature,
when the three-phase contact line is stabilized at the concave
regions and the liquid interface remains suspended between surface
asperities.   The stability of the Cassie-Baxter state could be
further improved in the case of hierarchical surface topography,
where nanoscale texture is imposed on microscale roughness, and it
might involve a combination of many concave and convex
segments~\cite{Noson16}. Recently, the equilibrium and stability of
Cassie-Baxter wetting states on microstructured surfaces were
analyzed by minimizing the surface free energy and the potential
energy of the external pressure~\cite{Lange06,Feng16}. An estimate
of the critical value of the external pressure required to overcome
the Laplace pressure due the curved interface was obtained for a
number of typical 3D microstructures~\cite{Lange06,Feng16}. In
general, however, the shape of the liquid interface and the critical
pressure for the transition from the Cassie-Baxter to Wenzel states
on structured surfaces have to be determined numerically.

\vskip 0.05in

In recent years, the wetting behavior of liquid droplets on
nanotextured surfaces was extensively studied using molecular
dynamics (MD)
simulations~\cite{Koishi09,Yong09,Yen12,Korea13,ZhaoSci13,Gendt14,ZhaoNano15,Das16,Ma16,Tsao16,JYan17}.
For instance, it was shown that the free-energy barrier associated
with the transition from the Wenzel-to-Cassie state is much higher
than from the Cassie-to-Wenzel state, provided that the height of
pillars at the hydrophobic surface is sufficiently
large~\cite{Koishi09}.  It was also found that the apparent contact
angle depends sensitively on the pillar cross-sectional shape and
height, spacing between pillars, crystal plane orientation at the
tops of the pillars, and the interaction energy between pillar atoms
and droplet
molecules~\cite{Koishi09,Korea13,Gendt14,Das16,Ma16,Tsao16}.
Interestingly, during lateral motion of a liquid droplet over
deformable pillar-arrayed substrate, the flexible hydrophilic
pillars can accelerate the liquid upon approach and pin the receding
contact line~\cite{ZhaoSci13}.  Furthermore, it was recently
demonstrated that the wetting transition of a water droplet at a
pillar-arrayed surface also depends on the charge density at the
base substrate~\cite{Das16}.    However, despite considerable
computational efforts, the atomic details of the wetting transition
at nanostructured surfaces with re-entrant surface curvature remain
not fully understood.

\vskip 0.05in

In this paper, molecular dynamics and continuum simulations are
preformed to investigate the effects of external pressure and
surface energy on the wetting transition at nanotextured surfaces
with re-entrant curvature.  We consider a polymeric liquid film
partially wetting the surface of a spherical particle, which is
fixed rigidly at a solid substrate.  It will be shown that the
position of the liquid/vapor interface is determined by the particle
radius and the local contact angle at the particle surface.  With
increasing external pressure on the liquid film, the liquid/vapor
interface is displaced towards the substrate and becomes more curved
until it touches the substrate at the critical pressure.  We find an
excellent agreement between the results of molecular dynamics
simulations and numerical minimization of the surface energy for the
critical pressure as a function of the local contact angle.

\vskip 0.05in

The rest of the paper is organized as follows. The description of
molecular dynamics and continuum simulations is given in the next
section.  The relationship between the local contact angle and the
surface energy for a liquid droplet on a flat substrate, and the
pressure dependence of the shape and location of the suspended
liquid film as well as the comparative analysis of the critical
pressure as a function of the contact angle are presented in
Sec.\,\ref{sec:Results}.   The results are summarized in the last
section.

\section{Simulation Details}
\label{sec:MD_Model}

\begin{figure}[t]
\includegraphics[width=14.cm,angle=0]{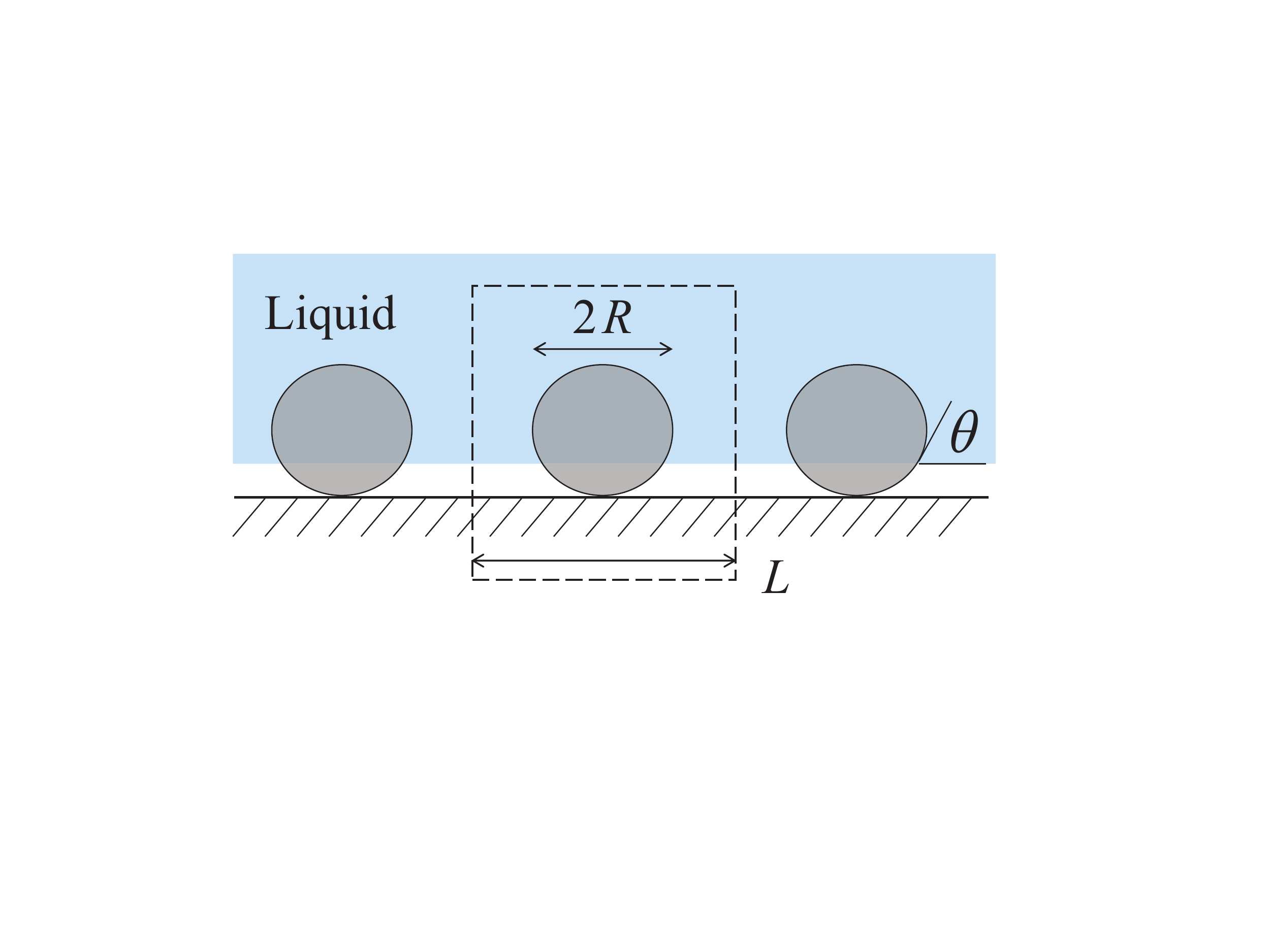}
\vskip -1.45in \caption{(Color online) A schematic illustration of
the structured interface that consists of an array of spherical
particles attached to the solid substrate and a suspended liquid
film.  The local contact angle of the liquid interface at the
surface of the solid particle is denoted by $\theta$. The dashed box
indicates the spatial domain used in molecular dynamics and
continuum simulations.}
\label{fig:schematic}
\end{figure}

We study wetting properties of structured interfaces that consist of
an array of spherical particles on a solid substrate and a suspended
liquid film, as shown schematically in Fig.\,\ref{fig:schematic}.
The large-scale molecular dynamics simulations were carried out
using the parallel code LAMMPS developed at Sandia National
Laboratories~\cite{Lammps}.  In our model, any two atoms interact
via the truncated Lennard-Jones (LJ) potential
\begin{equation}
V_{LJ}(r)=4\,\varepsilon\,\Big[\Big(\frac{\sigma}{r}\Big)^{12}\!-\Big(\frac{\sigma}{r}\Big)^{6}\,\Big],
\label{Eq:LJ}
\end{equation}
where the parameters $\varepsilon$ and $\sigma$ represent the energy
and length scales of the liquid phase. The interaction between atoms
of a liquid and a solid is also described by the LJ potential but
with the parameters $\varepsilon_{\rm wf}$ and $\sigma_{\rm wf}$
that are measured in units of $\varepsilon$ and $\sigma$,
respectively.   The solid atoms are fixed on either lattice sites or
on a surface of a sphere and they do not interact with each other.
Furthermore, the size of liquid and solid atoms is chosen to be the
same, i.e., $\sigma_{\rm wf} = \sigma$, throughout the study. For
computational efficiency, the cutoff radius is set
$r_{c}=2.5\,\sigma$ for all types of interactions.

\vskip 0.05in


We consider a polymeric fluid, where atoms are connected to form
linear chains with $N_p=10$ atoms each, which makes the surface
tension of the liquid/vapor interface greater than for monatomic
liquids~\cite{Grest03}. In addition to the LJ potential, the
interaction between nearest-neighbor atoms in a polymer chain is
described via the FENE (finitely extensible nonlinear elastic)
potential
\begin{equation}
V_{FENE}(r)=-\frac{k}{2}\,r_{\!o}^2\ln[1-r^2/r_{\!o}^2],
\end{equation}
with the parameters $k\,{=}\,30\,\varepsilon\sigma^{-2}$ and
$r_{\!o}\,{=}\,1.5\,\sigma$~\cite{Kremer90}.   The combination of
the LJ and FENE potentials with this parametrization yields an
effective harmonic potential that allows vibration of
nearest-neighbors but prevents polymer chains from unphysical
crossing each other~\cite{Kremer90}.  We also comment that
relatively short polymer chains considered in the present study form
a polymer melt well below the entanglement limit of about $70$ beads
per chain~\cite{Kremer90}.  The MD simulations were carried out at a
constant temperature of $1.0\,\varepsilon/k_B$, which was maintained
via the Nos\'{e}-Hoover thermostat applied to the fluid
phase~\cite{Lammps}. Here, $k_B$ denotes the Boltzmann constant. At
this temperature, the dependence of the surface tension at the
liquid/vapor interface as a function of the chain length was
reported in Ref.\,\cite{Grest03}. The equations of motion were
solved numerically using the velocity Verlet integration
algorithm~\cite{Allen87,Lammps} with the time step $\triangle
t_{MD}=0.005\,\tau$, where $\tau=\sigma\sqrt{m/\varepsilon}$ is the
characteristic LJ time.

\vskip 0.05in


Our computational domain includes a liquid film in contact with a
solid sphere, which is in turn rigidly fixed on a solid substrate
(see the dashed box in Fig.\,\ref{fig:schematic}).  The solid
substrate consists of $2500$ atoms arranged on square lattice sites
within the $xy$ plane with dimensions $L \times L = 50.0\,\sigma
\times 50.0\,\sigma$, and, therefore, the areal density is
$1.0\,\sigma^{-2}$.  The interaction energy between fluid monomers
and atoms of the lower stationary substrate is fixed to
$1.0\,\varepsilon$. Secondly, the solid particle consists of $4000$
atoms uniformly distributed on a surface of a sphere with the radius
$R=17.8\,\sigma$, and, correspondingly, with the areal density of
$1.0\,\sigma^{-2}$. Thus, the important input parameter that
controls wetting properties of the solid sphere is the LJ
interaction energy between its atoms and fluid monomers, i.e.,
$\varepsilon_{\rm wf}$.

\vskip 0.05in

Next, the fluid phase is composed of $85000$ monomers that
altogether form $8500$ polymer chains. Periodic boundary conditions
are imposed in the $\hat{x}$ and $\hat{y}$ directions parallel to
the stationary lower substrate.   In MD simulations, the fluid phase
is confined from above by the upper wall (not shown in
Fig.\,\ref{fig:schematic}), which also consists of $2500$ atoms that
are arranged on a square plane with the linear size
$L=50.0\,\sigma$. The LJ interaction energy between fluid monomers
and atoms of the upper wall is also set to $1.0\,\varepsilon$. In
contrast to the stationary substrate, the upper wall can move freely
under forces from the fluid atoms and under the external pressure
$P$, which is applied along the $\hat{z}$ direction toward the solid
substrate. Thus, within this geometry, the position of the liquid
film relative to the stationary substrate and shape of the
liquid-vapor interface are determined by two parameters, i.e.,
$\varepsilon_{\rm wf}$ and $P$. We finally comment that the effects
of gravity and entrapped gas were not considered in the present
study.

\vskip 0.05in


A complimentary analysis of the liquid film trapped by the surface
of a spherical particle was performed using the numerical software
Surface Evolver~\cite{Brakke91}. In the continuum simulations, the
liquid interface gradually evolves toward a state with a minimum
energy by the gradient descent method.   To facilitate the
comparison, the geometry of the problem shown in
Fig.\,\ref{fig:schematic} includes the same ratio of the sphere
radius and the linear size of the simulation cell, $R/L$, as the one
used in the MD setup.  As discussed in the next section, the results
of MD simulations for the contact angle as a function of the surface
energy were used as input parameters in the continuum analysis.

\section{Results}
\label{sec:Results}


At the atomic level, the local contact angle and shape of the
vapor-liquid interface near the contact line depend on densities of
liquid and solid phases and the interaction energy between fluid
monomers and wall
atoms~\cite{Davis99,Milchev02,Coninck09,Snoeijer11}.   In order to
determine the local contact angle as a function of the surface
energy, we first performed a set of separate MD simulations and
considered a polymeric droplet in contact with a solid substrate.
More specifically, the droplet was composed of 8500 bead-spring
polymer chains, each of length $N_p=10$, which were placed in
contact with a solid substrate that contains $25600$ atoms rigidly
fixed on a square lattice with lateral dimensions $160.0\,\sigma
\times 160.0\,\sigma$, and, thus, with the areal density of
$1.0\,\sigma^{-2}$. Note that the areal density of atoms on the
solid substrate is chosen to be the same as that for the solid
spherical particle described in the previous section.   In addition,
the thermostatting procedure, interaction potentials, and the
relative size of solid and fluid atoms were kept the same as in
Sec.\,\ref{sec:MD_Model}.

\vskip 0.05in

\begin{figure}[t]
\includegraphics[width=15.cm,angle=0]{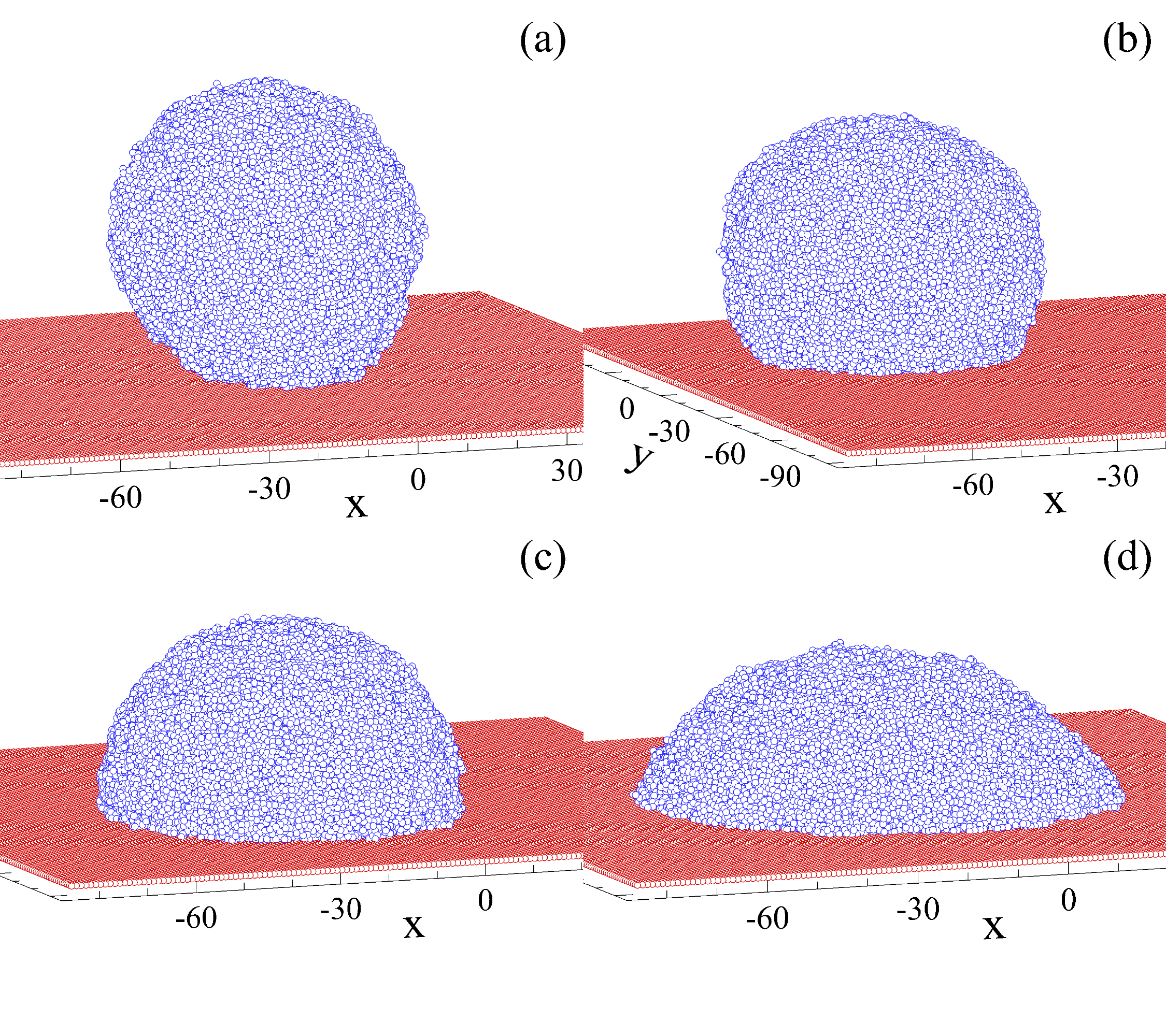}
\vskip -0.25in \caption{(Color online) Snapshots of liquid droplets
that consist of $85\,000$ atoms (blue circles) in contact with the
solid substrate (red circles) with the wall-fluid interaction energy
(a) $\varepsilon_{\rm wf}\!=\!0.2\,\varepsilon$, (b)
$\varepsilon_{\rm wf}\!=\!0.4\,\varepsilon$, (c) $\varepsilon_{\rm
wf}\!=\!0.6\,\varepsilon$, and (d) $\varepsilon_{\rm
wf}\!=\!0.8\,\varepsilon$. The dimensions of the solid substrate are
$160.0\,\sigma \times 160.0\,\sigma$. }
\label{fig:snapshot_droplets}
\end{figure}
\begin{figure}[t]
\includegraphics[width=12.cm,angle=0]{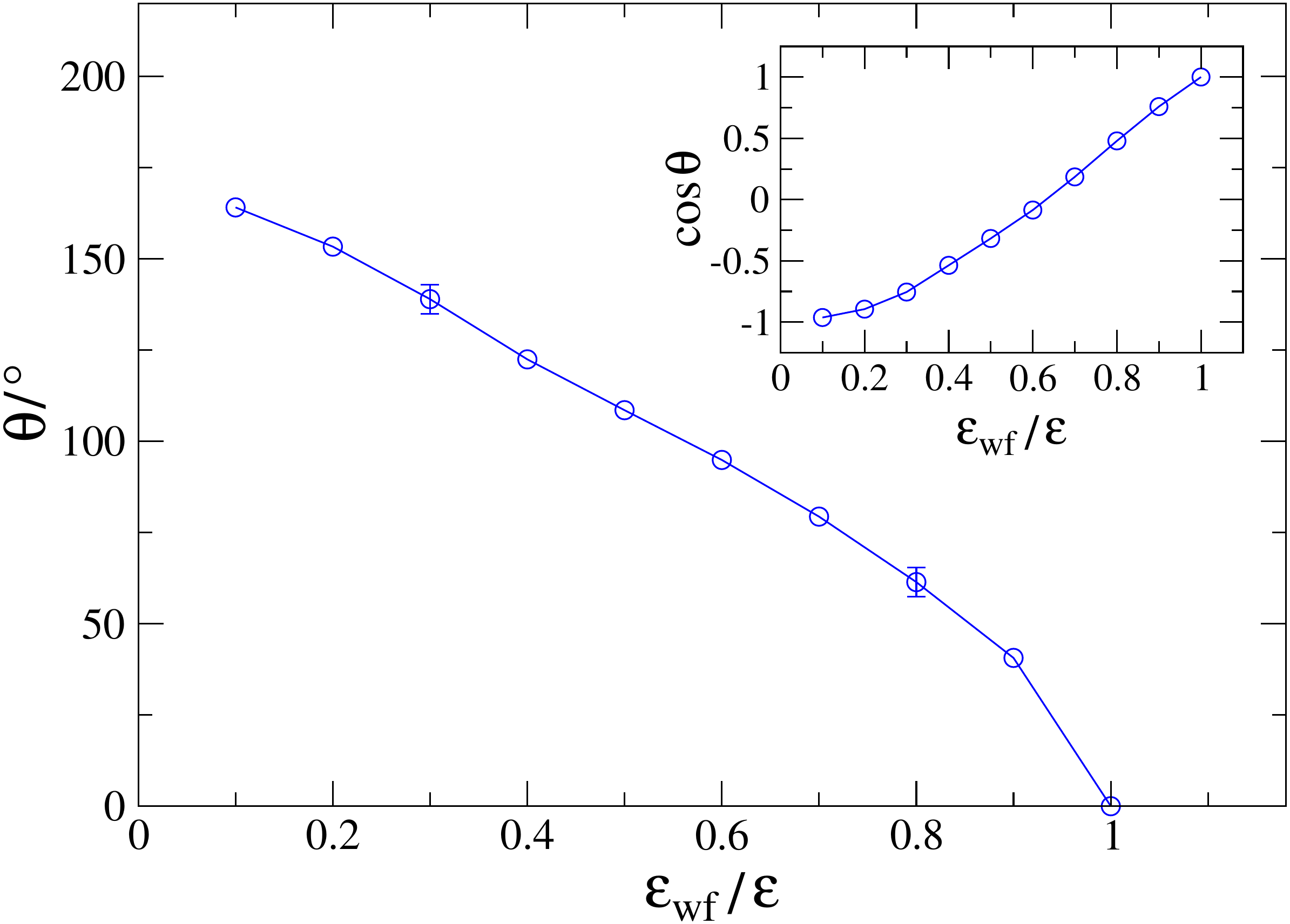}
\caption{(Color online) The dependence of the contact angle as a
function of the surface energy $\varepsilon_{\rm wf}/\varepsilon$
for liquid droplets residing on a crystalline substrate. The data
are obtained from the best fit of the liquid-vapor interface to a
spherical cap. The inset shows the same data replotted as
$\text{cos}\,\theta(\varepsilon_{\rm wf})$.}
\label{fig:contact_angle_eps}
\end{figure}

Typical examples of liquid droplets partially wetting solid
substrates are shown in Fig.\,\ref{fig:snapshot_droplets} for the
selected values of the wall-fluid interaction energy.   For each
value of $\varepsilon_{\rm wf}$, the system was equilibrated at the
temperature of $1.0\,\varepsilon/k_B$ for at least $3\times10^6$ MD
steps. It can be clearly observed that with increasing surface
energy, the shape of the liquid-vapor interface becomes less curved
and the liquid-solid contact area increases.   We comment that with
further increasing wall-fluid interaction energy, $\varepsilon_{\rm
wf} \geqslant 1.0\,\varepsilon$, the solid substrate becomes
uniformly covered with a flat liquid film with thickness of about
$3-4\,\sigma$ (not shown).    The contact angle was extracted by
fitting a spherical cap to a liquid droplet and then averaged over
several independent configurations for each value of
$\varepsilon_{\rm wf}$.    The summary of the data is shown in
Fig.\,\ref{fig:contact_angle_eps} for different the wall-fluid
interaction energies.  As expected, the local contact angle,
$\theta$, decreases monotonically from a large value of about
$164.1^{\circ}$, in the case of droplet on a nearly nonwetting
substrate to zero for a thin liquid film.   These values of the
local contact angle were used as input parameters to model the shape
of liquid interfaces around a spherical particle using the energy
minimization method implemented in the software Surface Evolver
(described below).

\vskip 0.05in

\begin{figure}[t]
\includegraphics[width=15.cm,angle=0]{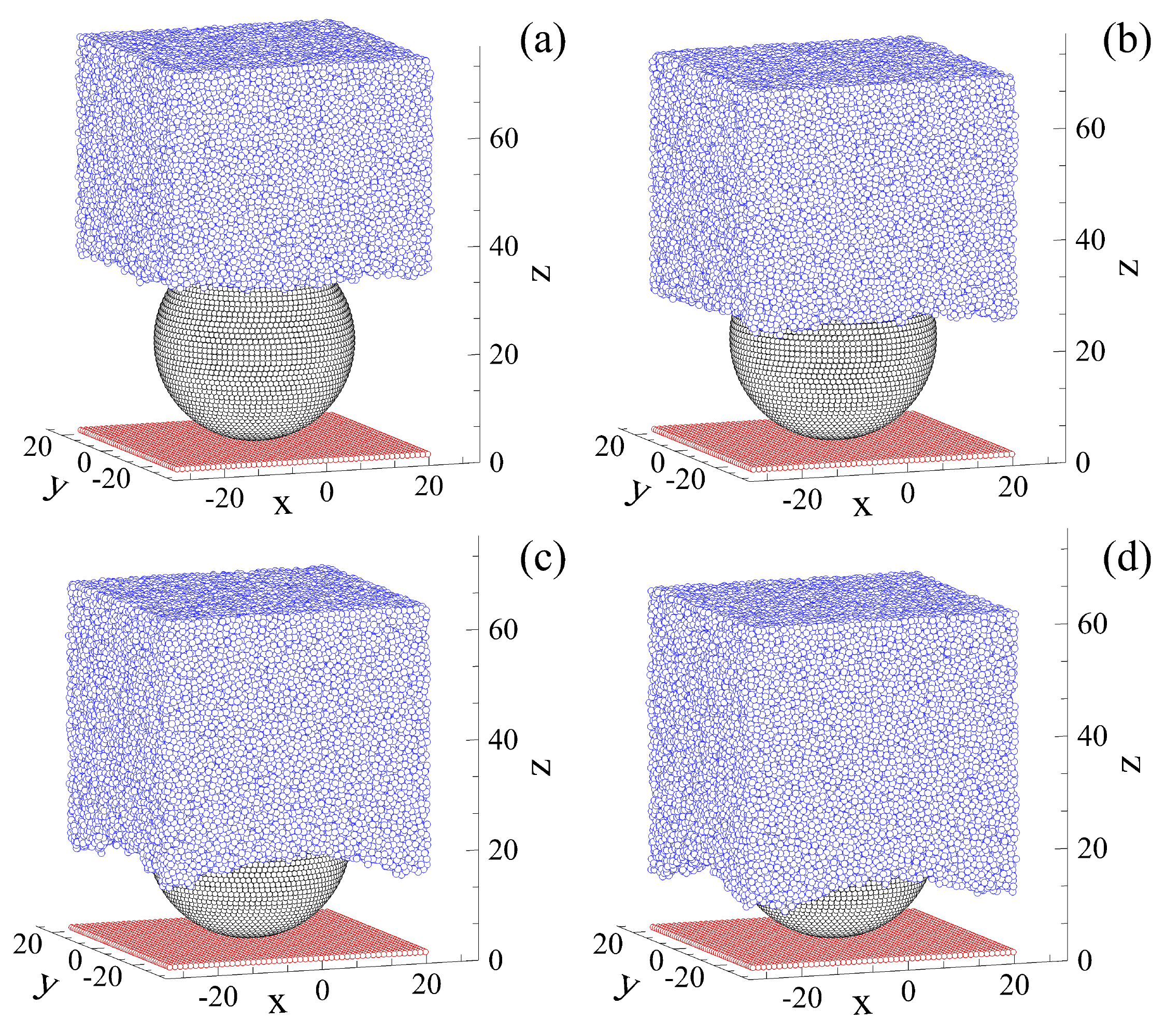}
\caption{(Color online) Snapshots of the liquid film suspended on
the spherical particle for the wall-fluid interaction energy
$\varepsilon_{\rm wf}\!=\!0.3\,\varepsilon$ and the vertical
pressure (a) $P=0$, (b) $P=0.02\,\varepsilon\,\sigma^{-3}$, (c)
$P=0.05\,\varepsilon\,\sigma^{-3}$, and (d)
$P=0.055\,\varepsilon\,\sigma^{-3}$.   The averaged value of the
local contact angle is $\theta=138.94^{\circ}$. }
\label{fig:film_sphere_eps03}
\end{figure}
\begin{figure}[t]
\includegraphics[width=15.cm,angle=0]{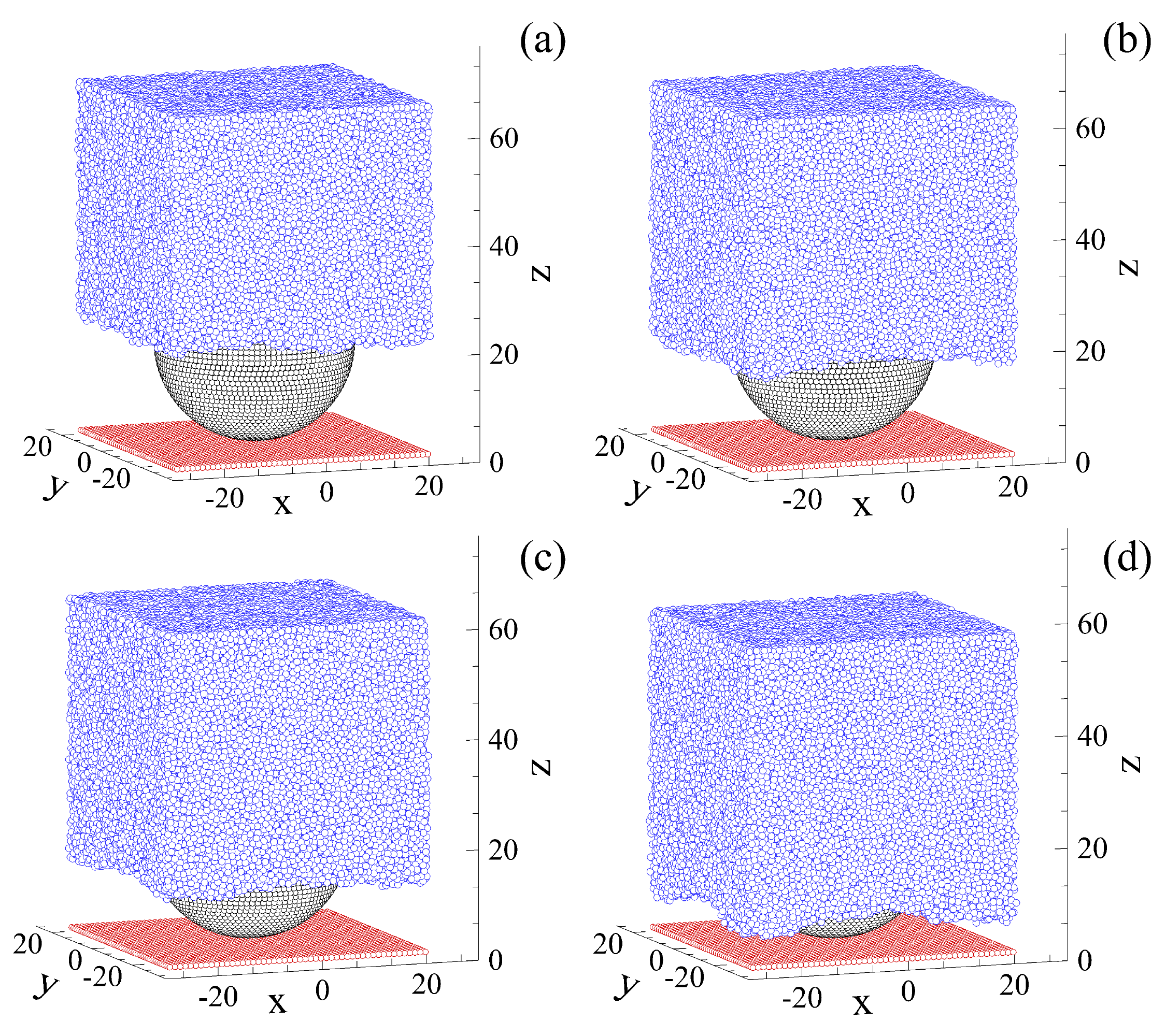}
\caption{(Color online) A liquid film partially wetting the
spherical particle for the wall-fluid interaction energy
$\varepsilon_{\rm wf}\!=\!0.6\,\varepsilon$ and the vertical
pressure (a) $P=0$, (b) $P=0.01\,\varepsilon\,\sigma^{-3}$, (c)
$P=0.02\,\varepsilon\,\sigma^{-3}$, and (d)
$P=0.03\,\varepsilon\,\sigma^{-3}$.   The local contact angle at the
particle surface is $\theta=94.86^{\circ}$. }
\label{fig:film_sphere_eps06}
\end{figure}

We now turn our discussion to the analysis of partially wetting
states on textured surfaces covered by spherical particles.   To
make the problem computationally feasible, only one spherical
particle with periodic boundary conditions in lateral directions was
considered, as illustrated in Fig.\,\ref{fig:schematic}.   The
spherical particle is attached rigidly to the solid substrate, whose
atoms interact with the fluid monomers with the LJ energy of
$1.0\,\varepsilon$.   Two series of snapshots extracted from MD
simulations are displayed in Fig.\,\ref{fig:film_sphere_eps03} for
$\varepsilon_{\rm wf}\!=\!0.3\,\varepsilon$ and in
Fig.\,\ref{fig:film_sphere_eps06} for $\varepsilon_{\rm
wf}\!=\!0.6\,\varepsilon$.   In each case, the external pressure was
applied on the upper wall (not shown) along the negative $\hat{z}$
direction.   The liquid film and its interface were allowed to relax
during the time interval of $5\times10^3\,\tau$ each time the
pressure was incremented by $0.05\,\varepsilon\,\sigma^{-3}$. Notice
the finite thickness of the liquid-vapor interface where some chain
segments temporarily move away due to thermal fluctuations.

\vskip 0.05in


At zero applied pressure, the liquid-vapor interface is flat and its
location relative to the lower substrate is determined by the local
contact angle, $\theta(\varepsilon_{\rm wf})$, at the surface of the
solid sphere [\,see Fig.\,\ref{fig:film_sphere_eps03}\,(a) and
Fig.\,\ref{fig:film_sphere_eps06}\,(a)\,]. We checked that this
local contact angle is the same as the contact angle measured for a
polymer droplet on a flat substrate.  According to the results in
Fig.\,\ref{fig:contact_angle_eps}, the local contact angle is larger
at lower surface energy, and, therefore, the liquid interface is
located further away from the solid substrate in the case
$\varepsilon_{\rm wf}\!=\!0.3\,\varepsilon$.   It is clearly seen in
Figs.\,\ref{fig:film_sphere_eps03} and \ref{fig:film_sphere_eps06}
that upon increasing external pressure, the liquid film is displaced
closer to the solid substrate and the liquid-vapor interface becomes
more curved, especially near the corners of the simulation cell that
are located further away from the spherical particle and its
periodic images.

\vskip 0.05in

\begin{figure}[t]
\includegraphics[width=12.cm,angle=0]{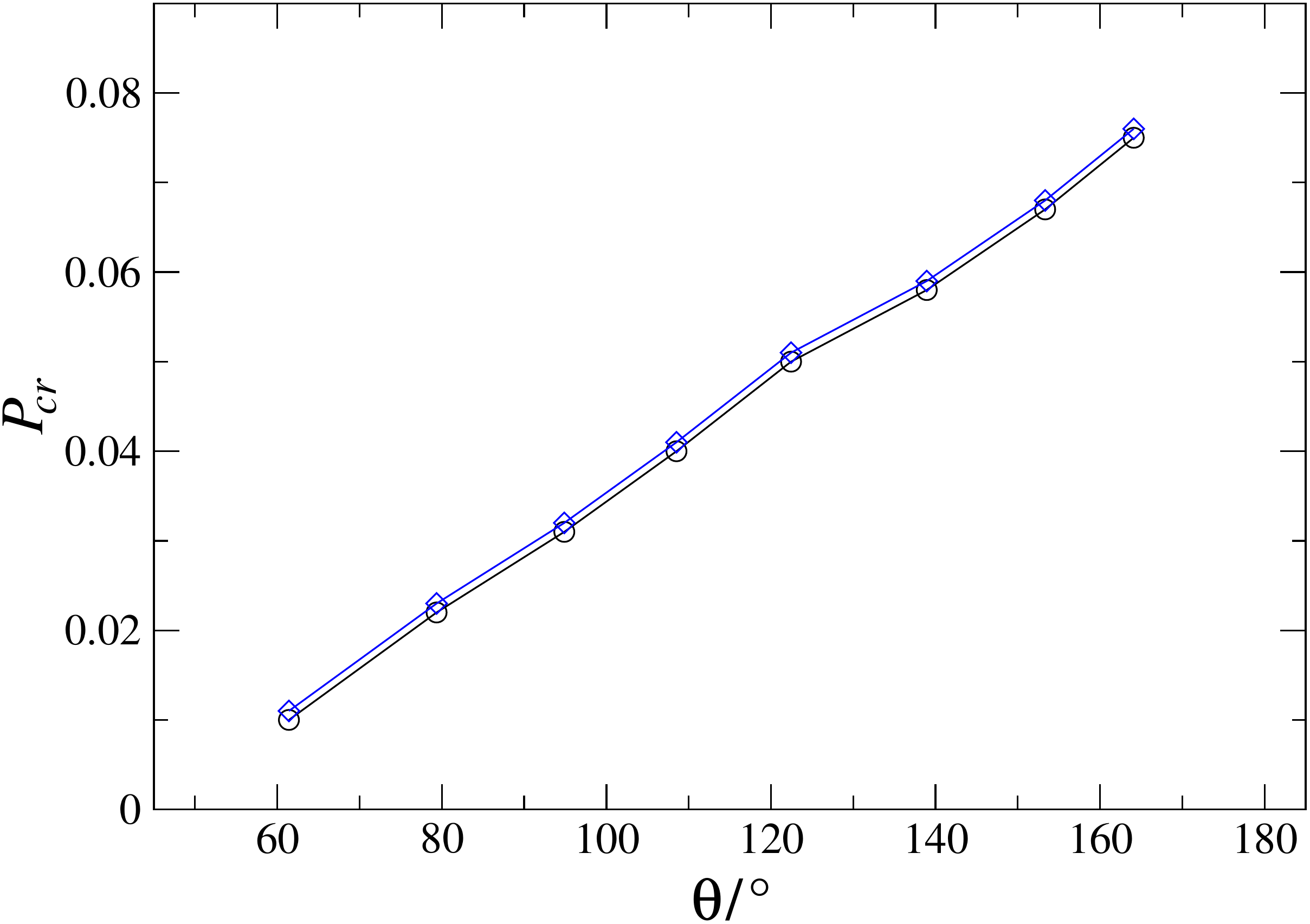}
\caption{(Color online) The critical pressure $P_{cr}$ (in units of
$\varepsilon\,\sigma^{-3}$) as a function of the local contact angle
$\theta$ (in degrees).  The largest external pressure at which the
liquid interface does not touch the lower substrate is denoted by
($\circ$). The threshold pressure associated with a wetting
transition to the Wenzel state is indicated by ($\diamond$). }
\label{fig:critical_pressure}
\end{figure}

We find that for each value of the wall-fluid interaction energy,
when the pressure is increased up to a critical value, the sagged
liquid/vapor interface touches the solid substrate, which triggers a
transition to a fully wetting Wenzel state.   Once the liquid film
completely wets the solid substrate and the spherical particle, it
remains in the wetting state even if the external pressure is
reduced.  In other words, the wetting transition is irreversible
upon decreasing pressure down to zero.   Furthermore, the variation
of the critical pressure as a function of the contact angle is
presented in Fig.\,\ref{fig:critical_pressure}.    To remind, the
values of the local contact angle at the surface of the spherical
particle were estimated from the shape of the interface of a liquid
droplet residing on a flat substrate, which has the same density as
the particle (see Fig.\,\ref{fig:contact_angle_eps}).  As evident in
Fig.\,\ref{fig:critical_pressure}, the critical pressure increases
for more nonwetting particles.   In the vicinity of the wetting
transition, the external pressure was incremented in smaller steps,
by $0.01\,\varepsilon\,\sigma^{-3}$, in order to resolve the
critical pressure more accurately.    Hence, for each value of
$\theta$, the maximum applied pressure at which the liquid interface
remains suspended at the particle surface is also included in
Fig.\,\ref{fig:critical_pressure}.


\vskip 0.05in


An alternative approach to estimate the critical pressure and the
shape of three-dimensional liquid interfaces at structured surfaces
involves an energy minimization implemented in the software Surface
Evolver~\cite{Brakke91}.   In this method, the surfaces of the lower
substrate, spherical particle, and liquid interface are represented
as a simplicial complex that consists of vertices, edges and facets.
Similar to the MD setup, periodic boundary conditions were applied
in the directions parallel to the solid substrate in order to mimic
a periodic array of spherical particles. Furthermore, the ratio of
the particle radius to the linear size of the lower substrate was
chosen to be the same as in MD simulations, $R/L=0.366$, where a
finite size of a liquid monomer was taken into account.   The liquid
film was initially placed above the spherical particle parallel to
the solid substrate, and the liquid interface was allowed to evolve
iteratively to the state with the lowest energy using the gradient
descent method.   As in MD simulations, the liquid volume is
sufficiently large so that the spherical particle can be completely
immersed into the liquid phase in the Wenzel state.   At each step
during the iteration procedure, the liquid interface is relaxed to
form the prescribed local contact angle with the surface of the
spherical particle. Note that the same values of the local contact
angle as reported in Fig.\,\ref{fig:contact_angle_eps} were used in
the continuum simulations.

\vskip 0.05in

\begin{figure}[t]
\begin{center}
\begin{tabular}{c c}
{\includegraphics[scale=0.4]{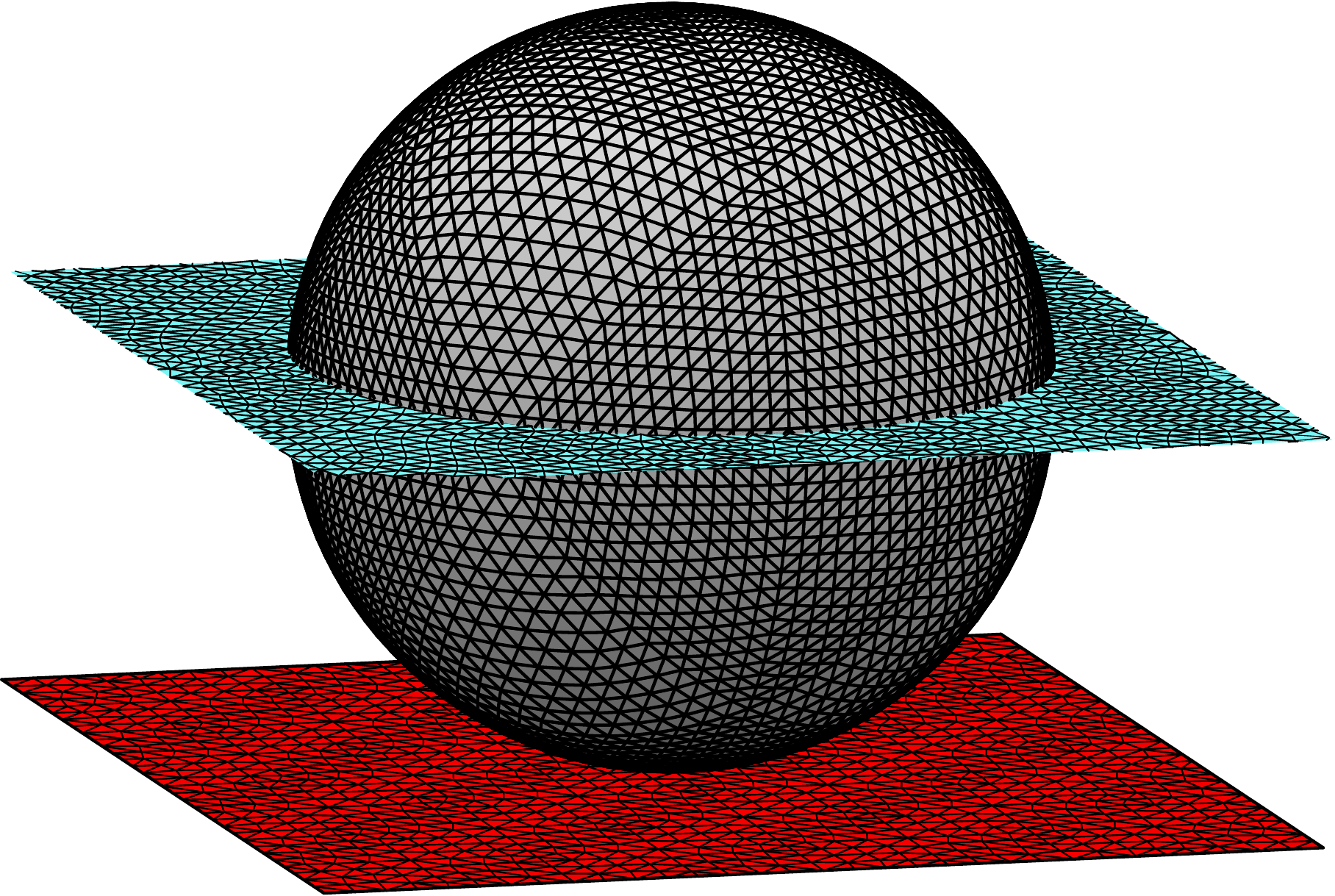} }  & {
\includegraphics[scale=0.4]{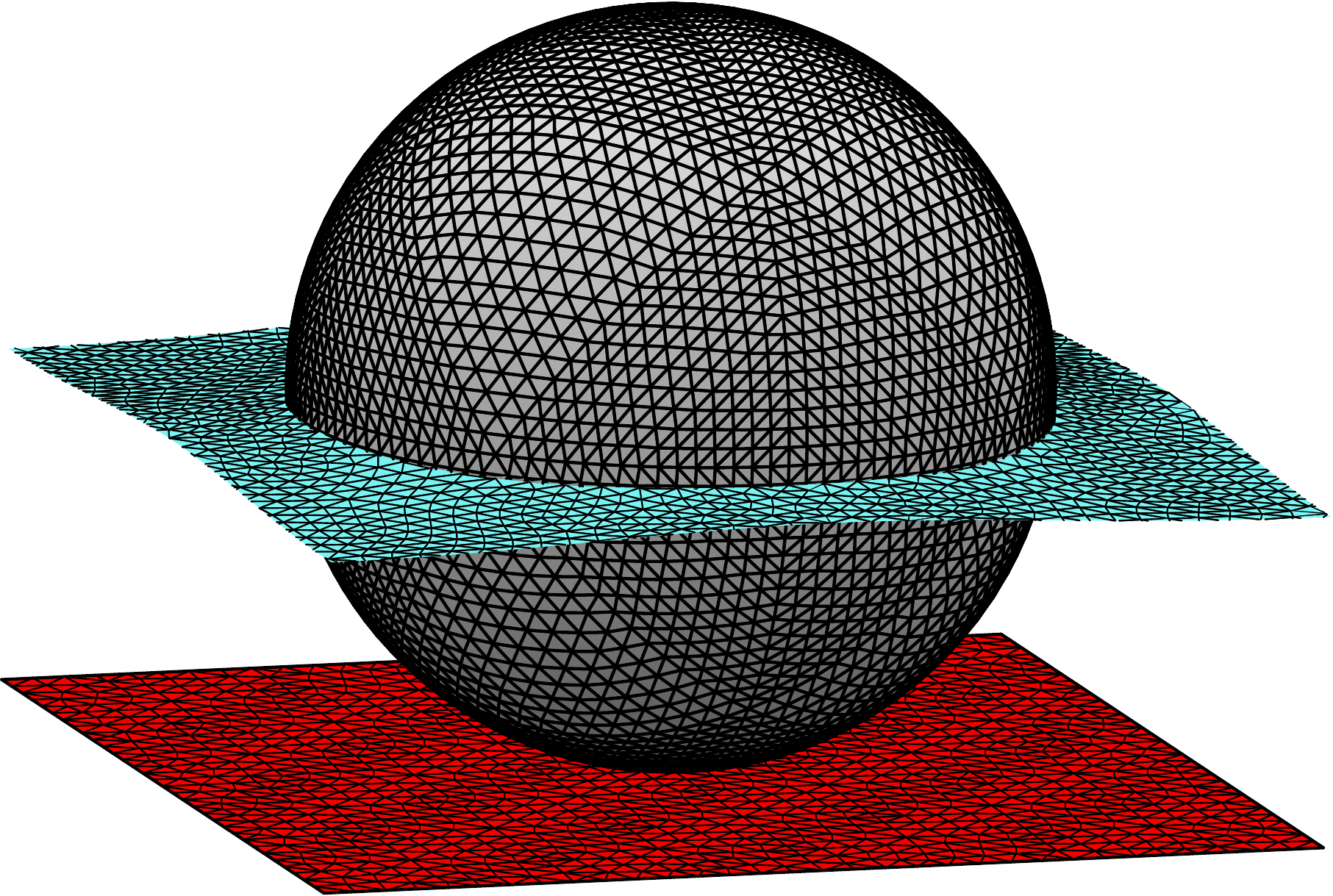} }
\\ {(a)} & {(b)} \\
\\
{ \includegraphics[scale=0.4]{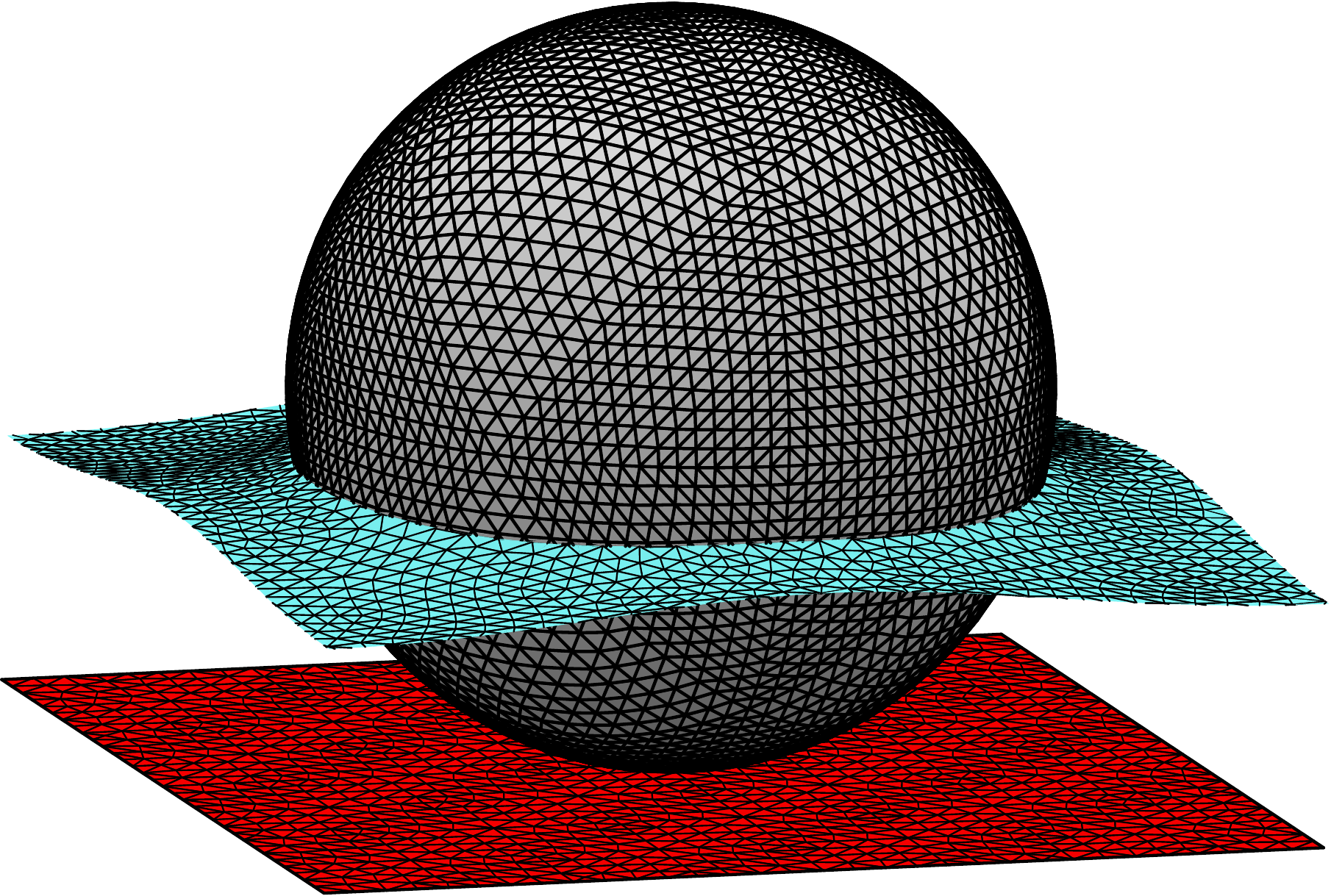}  } & {
\includegraphics[scale=0.4]{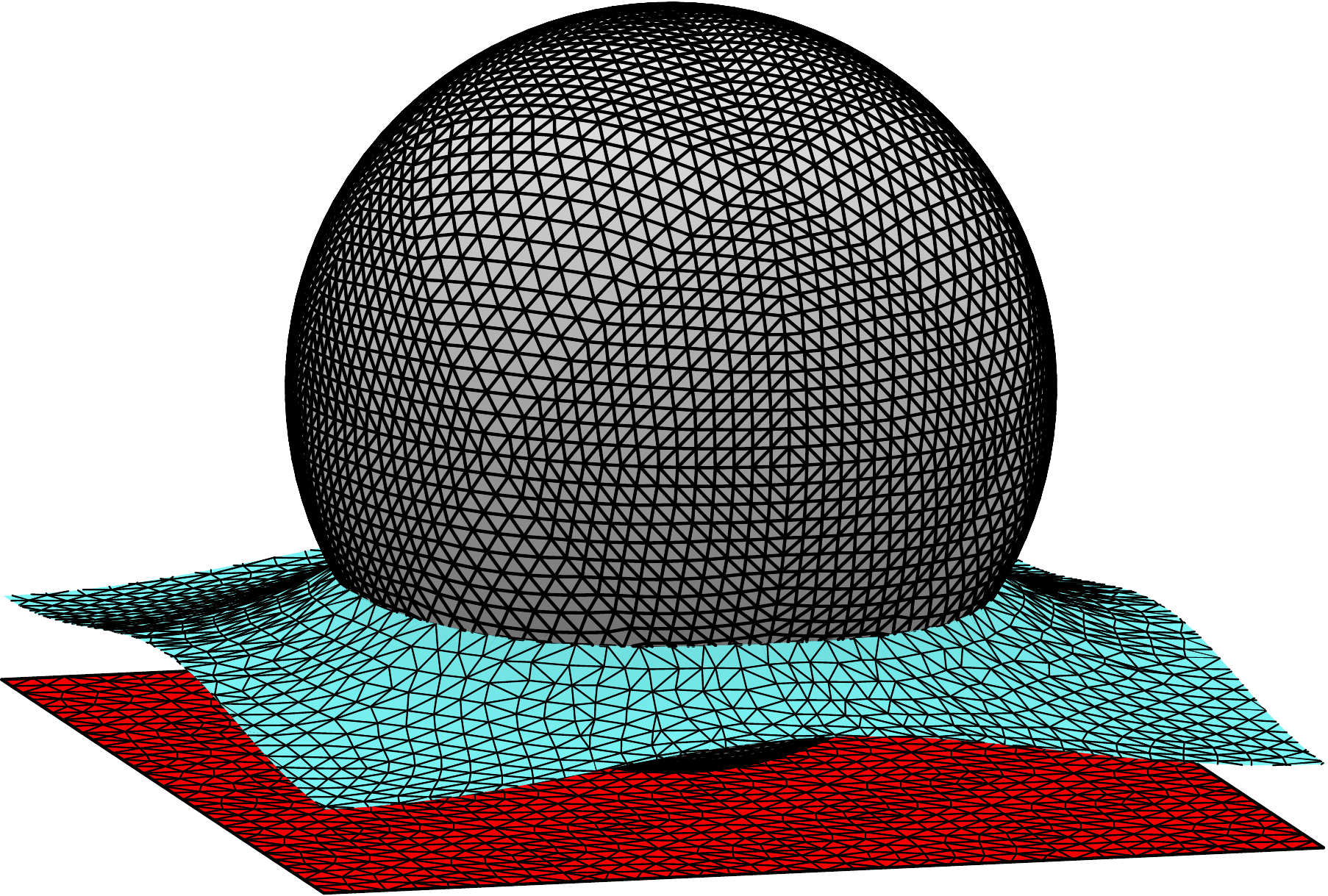}  }
\\ {(c)} & {(d)}
\end{tabular}
\end{center}
\vskip -0.4in \caption{(Color online) A sequence of snapshots of the
liquid/vapor interface obtained numerically using the software
Surface Evolver. The local contact angle at the surface of the solid
sphere is $\theta=94.86^{\circ}$ and the surface tension coefficient
is $\gamma=1.0\,\text{N/m}$.  The external pressure (a) $P=0$, (b)
$P=205\,\text{Pa}$, (c) $P=410\,\text{Pa}$, and (d)
$P=610\,\text{Pa}$. }
\label{fig:snapshots_surf_evol_th90}
\end{figure}
\begin{figure}[t]
\begin{center}
\begin{tabular}{c c}
{\includegraphics[scale=0.4]{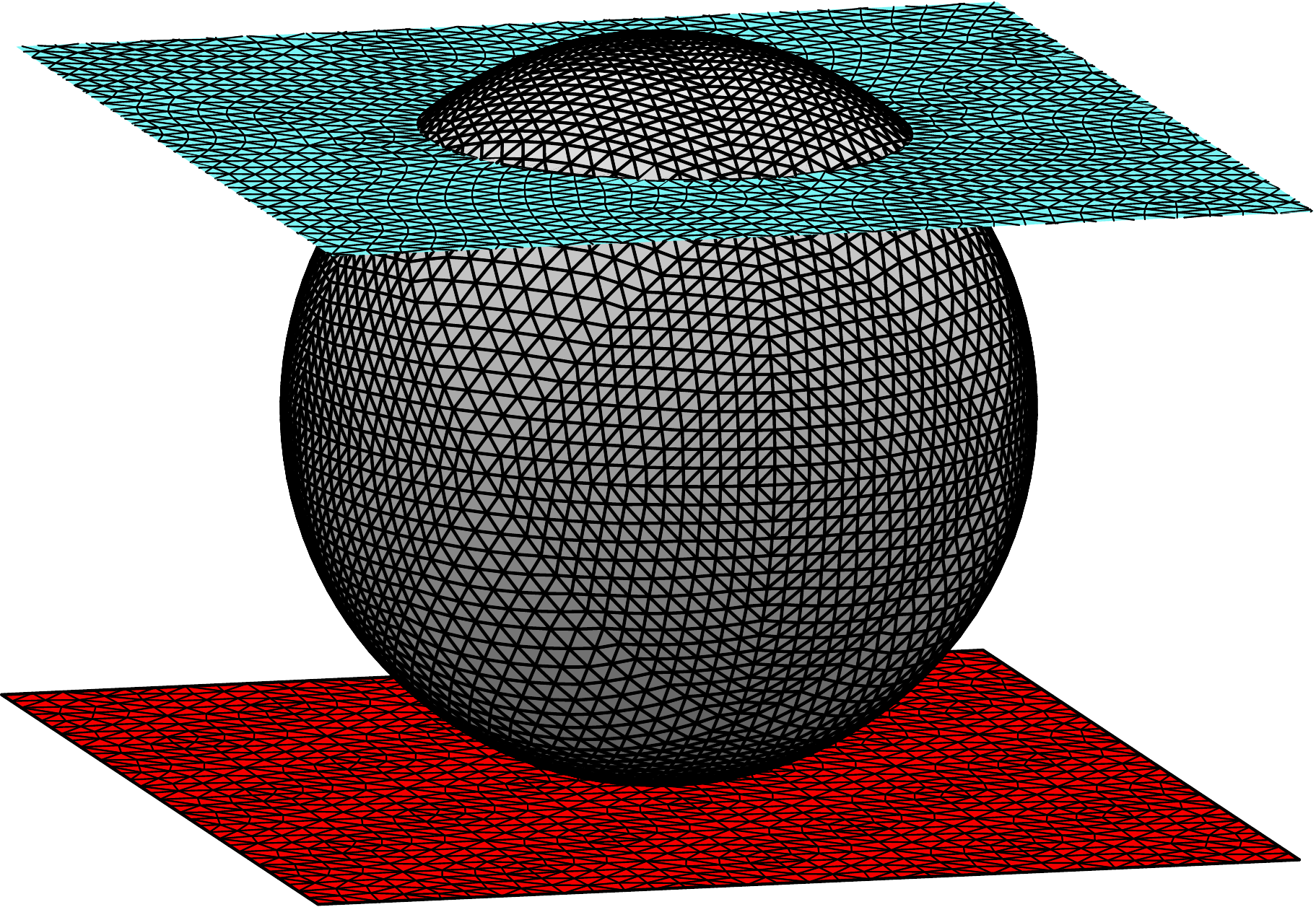}  }  & {
\includegraphics[scale=0.4]{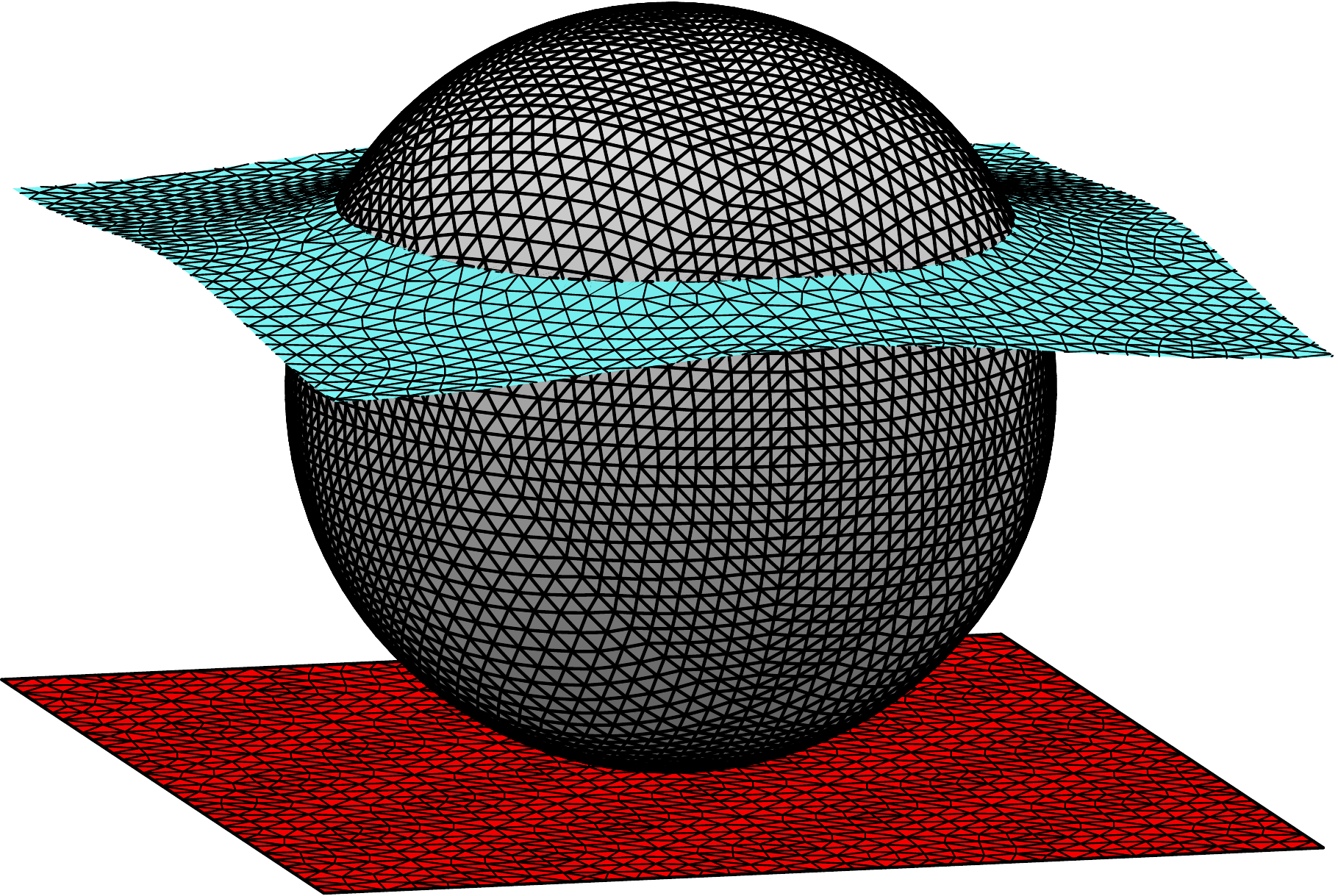} }
\\ {(a)} & {(b)}  \\
\\
{ \includegraphics[scale=0.4]{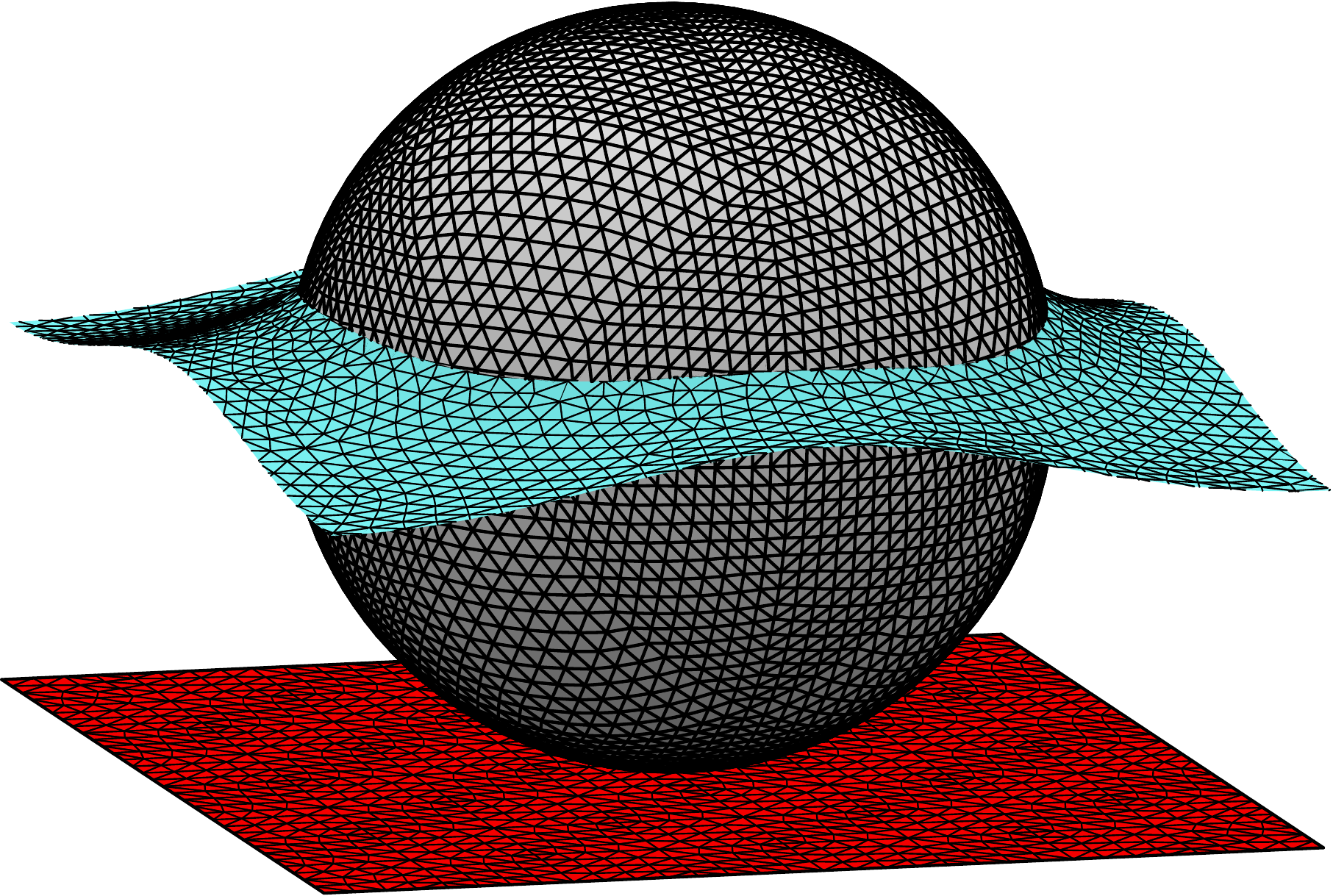}  } & {
\includegraphics[scale=0.4]{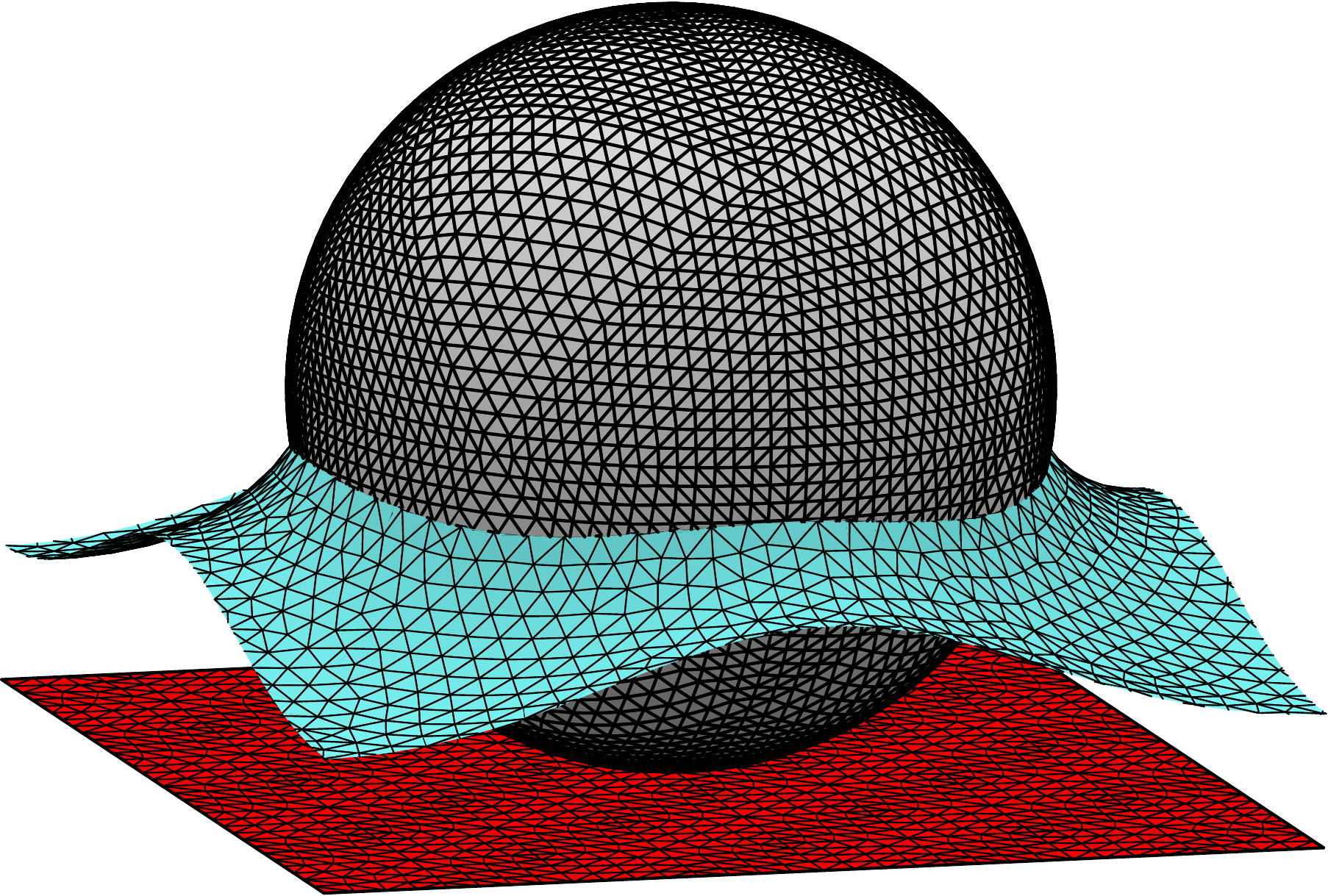}  }
\\ {(c)} & {(d)}
\end{tabular}
\end{center}
\vskip -0.4in \caption{(Color online) The shape of liquid/vapor
interfaces for the external pressures (a) $P=0$, (b)
$P=400\,\text{Pa}$, (c) $P=800\,\text{Pa}$, and (d)
$P=1200\,\text{Pa}$. The contact angle at the surface of the
spherical particle is $\theta=138.94^{\circ}$ and the surface
tension of the liquid/vapor interface is $\gamma=1.0\,\text{N/m}$. }
\label{fig:snapshots_surf_evol_th130}
\end{figure}

Typical snapshots of liquid/vapor interfaces around the spherical
particle are shown in Fig.\,\ref{fig:snapshots_surf_evol_th90} for
$\theta=94.86^{\circ}$ and in
Fig.\,\ref{fig:snapshots_surf_evol_th130} for
$\theta=138.94^{\circ}$ for the indicated values of the applied
pressure.   It can be observed that the liquid interface is flat in
the absence of the applied pressure, and it becomes more curved and
displaced towards the solid substrate at higher pressures.  Upon
further increasing external pressure, the liquid first touches the
solid substrate at the corners of the simulation cell and then
spreads at the substrate until the interface forms a prescribe
contact angle with the solid wall (not shown).   The critical
pressure of the wetting transition was determined with the accuracy
of $0.5\,\text{Pa}$.

\vskip 0.05in

\begin{figure}[t]
\includegraphics[width=12.cm,angle=0]{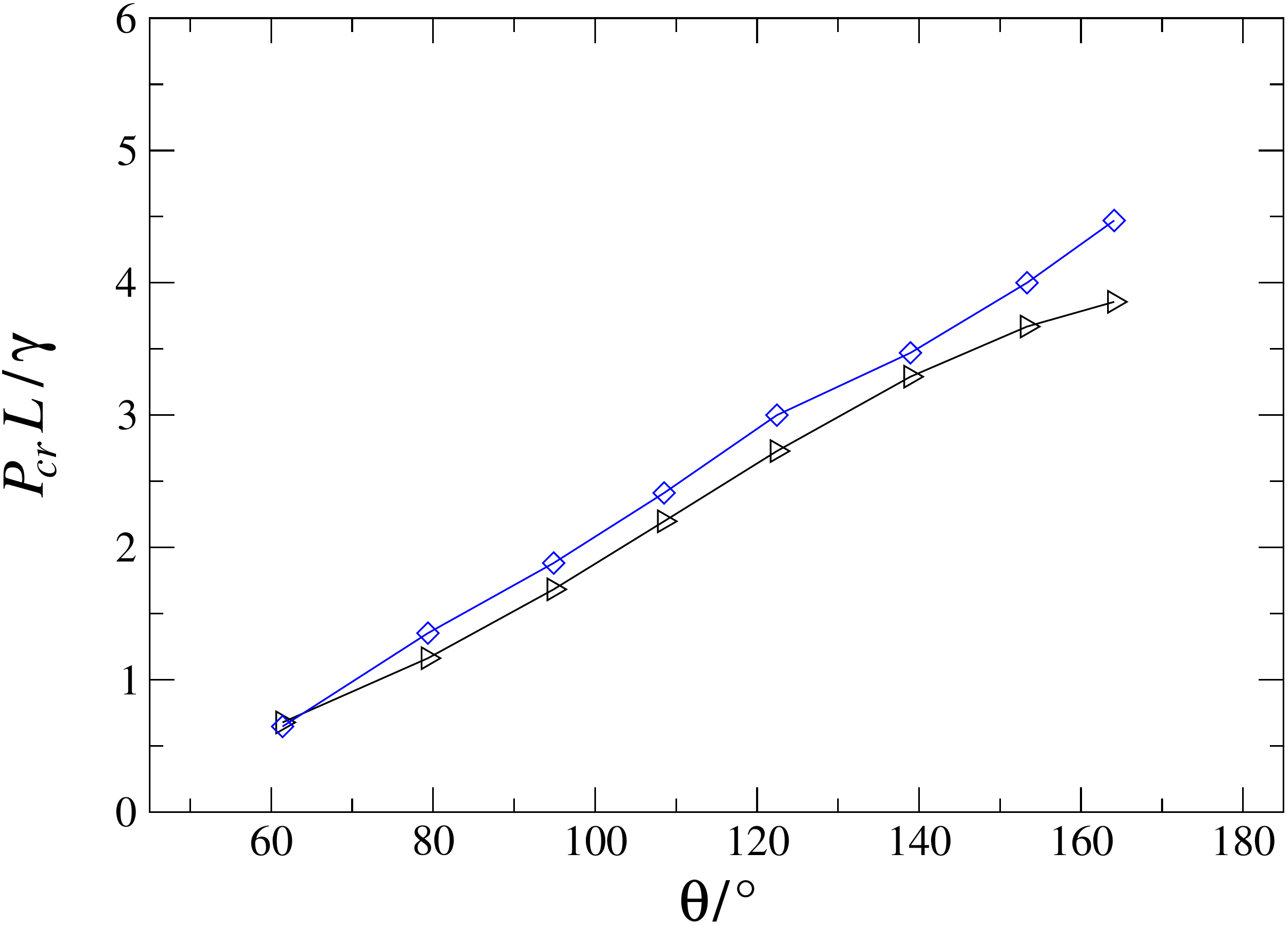}
\caption{(Color online) The dimensionless variable $P_{cr} L /
\gamma$ as a function of the local contact angle $\theta$. The MD
data are indicated by ($\diamond$) and continuum results are denoted
by ($\triangleright$).   Here, $L$ is the linear size of the lower
substrate and $\gamma$ is the surface tension of the liquid/vapor
interface.   The error bars are about the symbol size. }
\label{fig:pcr_comp_MD_cont}
\end{figure}

The results for the critical pressure for different contact angles
obtained from MD simulations and energy minimization are summarized
in Fig.\,\ref{fig:pcr_comp_MD_cont}.   In order to compare the
simulation results, we used the dimensionless variable $P_{cr} L /
\gamma$, where $L$ is the linear size of the solid substrate and
$\gamma$ is the surface tension coefficient.   In the Surface
Evolver, the input parameters were explicitly set to
$L=2.73\,\text{mm}$ and $\gamma=1.0\,\text{N/m}$.   By contrast, in
MD simulations the local contact angles and the surface tension are
determined by the LJ interaction parameters and need to be measured
separately.   In the previous study, the surface tension was
estimated to be $\gamma = 0.85\,\varepsilon/\sigma^2$ for a thin
liquid film that consists of bead-spring 10-mers at
$T=1.0\,\varepsilon/k_B$ and zero ambient pressure~\cite{Grest03}.

\vskip 0.05in


It can be seen in Fig.\,\ref{fig:pcr_comp_MD_cont} that the critical
pressure obtained from MD simulations agree well with the continuum
predictions.   These results imply that a continuum stability
analysis of partially wetting states on nanotextured surfaces might
hold at length scales of about a few nanometers.   Thus, in our
setup the closest distance between the surface of the spherical
particle and its images is about 30 molecular diameters.  It should
be also mentioned that one contributing factor to the slight
discrepancy between two approaches involves a finite cutoff radius
of the LJ interaction potential.   In other words, if the sagged
liquid/vapor interface is located within the cutoff radius from the
solid substrate, then a spontaneous wetting of the substrate might
occur due to thermal fluctuations. We finally comment that only one
value of the areal fraction was considered in the present study,
i.e., $\phi_{S} = \pi R^2 / L^2 = 0.42$.  It is expected that with
decreasing areal fraction, the critical pressure is reduced due to
smaller curvature of the liquid/vapor interface at the contact with
the solid substrate~\cite{McKinley08}.

\section{Conclusions}

In summary, molecular dynamics simulations and numerical
minimization of the interfacial energy were carried out to study
wetting properties of the composite interface that consists of a
periodic array of spherical nanoparticles rigidly attached to a
solid substrate. The wettability of spherical particles was
controlled via the strength of the fluid-solid interaction energy,
which in turn determines the local contact angle.   It was shown
that at zero applied pressure, the liquid film is suspended at the
surface of solid particles and the distance to the solid substrate
is determined by the particle radius and the local contact angle.
Upon increasing external pressure, the liquid film is displaced
closer to the solid substrate but remains stable up to a critical
value due to re-entrant curvature of the particle surface.   It was
found that the shape of liquid interfaces and the critical pressure
of permeation to the solid substrate obtained from atomistic
simulations agree well with the results of the numerical
minimization of the interfacial energy. These results are important
for modeling partially wetting states on hierarchically textured
surfaces that contain surface roughness on multiple length scales.

\section*{Acknowledgments}


Financial support from the National Science Foundation (CNS-1531923)
is gratefully acknowledged.  The molecular dynamics simulations were
carried out using the LAMMPS numerical code~\cite{Lammps}.  The
analysis of liquid interfaces was performed using publicly available
computer program \textit{Surface Evolver} developed by
Prof.~K.~A.~Brakke at Susquehanna University.  Computational work in
support of this research was performed at Michigan State
University's High Performance Computing Facility and the Ohio
Supercomputer Center.



\bibliographystyle{prsty}

\end{document}